\documentclass[prb,showpacs,twocolumn]{revtex4}

   \usepackage{times}
   \usepackage{epsfig}

\begin{document}

   \title{\bf Kinetic model of II-VI(001) semiconductor surfaces: Growth rates in atomic layer epitaxy}

   \author{T. Volkmann}
   \author{M. Ahr}
   \author{M. Biehl}
   \affiliation{
     Institut f\"ur Theoretische Physik und Astrophysik,
     Sonderforschungsbereich 410,
     Julius-Maxmilians-Universit\"at W\"urzburg,
     Am Hubland, D-97074 W\"urzburg, Germany
   }

   \date{\today}

   \begin{abstract}
   We present a zinc-blende 
   lattice gas model of II-VI(001) surfaces,
   which is investigated by means of Kinetic Monte Carlo (KMC) simulations.  
   Anisotropic effective interactions between surface metal atoms
   allow for the description of, e.g.,
   the sublimation of CdTe(001), including the
   reconstruction of Cd-terminated surfaces and its
   dependence on the substrate temperature $T$.
   Our model also includes Te-dimerization and the potential presence of
   excess Te in a reservoir of weakly bound atoms at the surface. 
   We study the self-regulation of atomic layer epitaxy (ALE)
   and demonstrate how the interplay of the reservoir occupation
   with the surface kinetics results in two different regimes: 
   at high $T$ the growth rate is limited
   to $0.5$ layers per ALE cycle, whereas at low enough
   $T$ each cycle adds a complete layer of CdTe. 
   The transition between the two regimes occurs at 
   a characteristic temperature and its dependence on external 
   parameters is studied. Comparing the temperature dependence of the
   ALE growth rate in our model with experimental results for CdTe we
   find qualitative agreement.
   \end{abstract}    


   \maketitle

%
%
   
    \section{Introduction}  \label{introduction}

    Molecular beam epitaxy (MBE) and its variations continue to attract
    significant interest as a technique for the production of, for
    instance, high quality semiconductor films. In addition to its practical
    relevance, epitaxial growth is highly attractive from a theoretical
    point of view. It offers many challenging open questions and 
    provides a workshop in which to develop novel tools for the modeling
    and simulation of non-equilibrium systems in general. 
    Reviews of the experimental techniques and theoretical investigations
    can be found in, e.g., Refs. \onlinecite{pimpinelli,corfu}.  

    II-VI semiconductors, as a promising class of materials, have been the
    subject of many experimental studies, see Ref. \onlinecite{zwosechs} for a recent overview.
    Among other aspects, competing surface reconstructions
    and their interplay with growth or sublimation have been investigated.
    \cite{tatarenko,neureiter,faschinger,hartmann,cibert,tatarenko2}
    As an example we will concentrate on the CdTe(001) surface, in the following,
    but many of the results should apply to other II-VI compounds as well.
    The focus will be on atomic layer epitaxy (ALE), a technique 
    which aims at self-regulated layered growth by alternating deposition of 
    the elements in ultra high vacuum.  
    Specific properties of the material system and the attractive clarity 
    of the ALE technique allow for an efficient theoretical modeling of the 
    growth scenario. 
    
    In earlier investigations we employed two-dimensional lattice gas models
    of the CdTe(001) surface \cite{epl,flat} for the  investigation of the
    temperature and flux dependence of surface reconstructions.
    Chemical bonds and the influence
    of the underlying crystal were  accounted for by effective pairwise
    anisotropic interactions.
    Sublimation and ALE growth  have been studied in the framework of a
    simple cubic model,\cite{surflett} which accounted only for the most essential
    features of CdTe.
    The correct zinc-blende lattice was first considered in Ref. \onlinecite{aachen}, but here
    we extend the model significantly by taking into account
    the formation of Te-dimers at the surface. 
    More importantly, we will consider weakly bound excess Te, which
    provides a reservoir of highly mobile atoms. A similar approach with
    weakly bound arsenic has recently been suggested by
    Itoh et al.\cite{itoh98,itoh01} in order to describe the
    growth on reconstructed GaAs(001) and InAs(001) surfaces.

    Within the framework of our model we demonstrate how the surface
    reconstruction and limited Cd-coverage restrict the
    ALE growth rate at high temperatures.  Our model shows how
    this limitation is overcome if weakly bound excess Te is present 
    on the surface at low temperatures. We observe a sudden 
    increase of the growth rate with decreasing temperature, 
    in qualitative agreement with experiments.
    
    Due to the increased number of parameters compared with our earlier models
    a quantitative match with experimental observations is hard 
    to achieve at the present stage of the model. One possibility to
    overcome this problem is to combine density functional calculations
    (DFT) with Kinetic Monte Carlo (KMC) simulations as was done e.g. by 
    Grosse et al.\cite{grosseprb,grosseprl} in the context of growth on III-V(001) 
    semiconductor surfaces.
    However, the available DFT input for II-VI(001) surfaces, especially CdTe(001),
    does not suffice for a full parametrization of our growth model.

    The paper is organized as follows: before introducing the model and method
    in section \ref{model}, we summarize essential properties and
    experimental findings concerning II-VI(001) surfaces in the next section.
    In section \ref{results} we discuss the  simulated growth  scenario
    and  present the results.  The last section summarizes and gives an
    outlook on perspective projects based on this work.

%
%

    \section{Properties of the material system} \label{properties}

    A detailed overview of II-VI(001) surfaces, with emphasis on 
    Te-compounds, can be found in Refs. \onlinecite{tatarenko,cibert,tatarenko2}, for instance.
    Whereas we will mostly refer to CdTe,  many of the
    features mentioned here apply to other compounds, e.g. ZnTe or ZnSe.

    CdTe crystallizes in a zinc-blende lattice and consists
    of alternating layers of Cd and Te which are parallel to
    the (001) surface of the crystal.
    In the bulk, each of these layers forms a regular square lattice,
    cf. Figure \ref{sosbild}.  Cd atoms are bound to two neighboring Te atoms
    in the layer below with bonds oriented along the $[1\overline{1}0]$ direction,
    which will be termed $y$-direction for short. 
    Accordingly, Te is always bound to two underlying Cd along the $[110]$ or $x$-direction. 
  
    Under vacuum conditions and in absence of particle deposition, 
    the CdTe(001) surface is  terminated by a Cd-layer with limited
    coverage $\rho^{\rm Cd} \leq 1/2$.  At low temperatures $T$
    one finds a dominant $c(2\times2)$ vacancy structure, which 
    corresponds to a checkerboard like occupation of the available 
    square lattice sites. Above $T\approx 570 K$ the surface is dominated by a local $(2\times1)$
    ordering where the Cd atoms arrange in rows along the $[110]$ 
    or $x$-direction  which alternate with empty rows. 
    Despite the fact that significant sublimation sets in 
    even below $570 K$, the temperature driven
    re-ordering of the Cd-terminated surface has been discussed
    in terms of  a phase transition  in effective equilibrium.
    \cite{neureiter,cibert,epl,flat}

    Electron counting rules,\cite{pashley} 
    and density functional theory (DFT) calculations \cite{garcia,park,gundel}
    show that the simultaneous occupation of
    nearest neighbor (NN) sites in the $y$-direction should be extremely
    unlikely within a terminating Cd layer. Such pairs of NN-Cd
    would lead to an unfavorable local concentration of positive charges. 
    Note that in both the above mentioned arrangements 
    NN pairs of Cd-atoms in $y$-direction do not occur, indeed. 
    DFT calculations show furthermore that the 
    $c(2\times2)$ ordering has a slightly lower surface energy
    than the competing $(2\times1)$ structure. \cite{garcia,park,gundel}

    The $c(2\times2)$ ordered Cd-termination can be 
    restored in the presence of a stationary Cd flux even at high $T$.\cite{tatarenko,cibert}
    On the other hand, a steady  deposition of Te stabilizes Te-terminated surfaces.\cite{tatarenko,cibert}
    At relatively small fluxes and high temperatures
    a coverage $\rho^{\rm Te}\approx 1$ is observed,
    hence there is no vacancy structure as for Cd-terminated surfaces. 
    The Te atoms arrange in rows of dimers along the $[110]$-direction
    at the surface, corresponding to a $(2\times1)$ reconstruction pattern. 

    At high flux and low temperatures  
    Te-coverages $\rho^{\rm Te}\approx 3/2$ with a $(2\times1)$
    surface symmetry are observed. 
    It has been hypothesized
    that Te atoms form trimers,\cite{tatarenko,cibert}
    but the precise mechanism of binding the excess Te to the surface
   remains  unclear.\cite{tatarenko2}
  The evaporation of terminating Te-layers after ending deposition has been studied 
  in the different temperature regimes.\cite{tatarenko,cibert,tatarenko2}
  The analysis of the corresponding Arrhenius-plots supports the hypothesis
  of a very weak binding of excess Te.

  As we will demonstrate in the following, the influence of the 
  excess Te is of particular importance for  ALE growth. 
  One ALE cycle, i.e.\ a sequence of one Cd and one Te pulse, 
  could add a complete layer of CdTe to the system, in principle.
  However, in a wide range of temperatures, one observes a growth 
  rate of at most $1/2$ layer per ALE cycle.\cite{faschinger,hartmann}
  This limitation reflects the restricted
  coverage of Cd-terminated surfaces.
  Only for sufficiently low $T$ one finds the 
  expected rate of approximately 1 layer per cycle,
  with a temperature driven, sudden transition from one regime to the other.
  \cite{faschinger,hartmann} 
  We will explain within the framework of our
  model how the  above mentioned presence of weakly bound Te atoms at the
  surface allows to overcome the limitation. 

%
%

  \section{The model}  \label{model}

  \subsection{Representation of the crystal and energetics} 
  \begin{figure}[t]
    \begin{center}
    \includegraphics[scale=0.55]{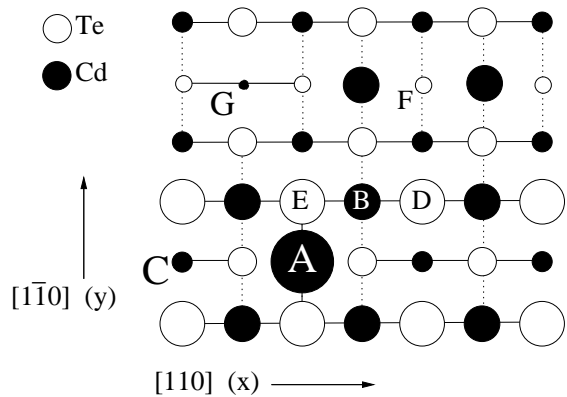} 
    \caption{\label{sosbild}
    Solid-On-Solid representation of the zinc-blende lattice. Filled / empty circles represent
    Cd and Te atoms, respectively, their size corresponds to the height above the substrate.
    The figure shows, for instance, a single Cd atom (A) on top of a Te-terminated terrace
    which extends along the $[110]$ direction. 
    Dotted lines represent chemical bonds of Cd to Te atoms in the layer underneath, 
    solid lines link Te with underlying Cd. Only bonds between visible atoms are displayed. 
    Atoms A and D are examples of mobile particles, whereas
    B and E have more than two chemical bonds and are considered immobile. Sites C and G
    are available for Cd-adsorption, whereas F can be occupied by a Te atom.
    Particle A could hop, for example, to site C in a diffusion move, or onto G
    descending the terrace edge. As an example for Te diffusion, D could hop to site F.}
    \end{center}
  \end{figure}

    As a simplification we exclude the incorporation of defects or vacancies  
    into the crystal. Hence, the bulk can be represented in a Solid-On-Solid (SOS) manner, 
    cf. Figure \ref{sosbild}. 
    Any atom with more than two chemical binding partners, i.e.\ with at least one 
    NN in the layer above, is considered immobile. When measuring the surface coverage $\rho^{\rm Cd/Te}$
    with Cd/Te atoms we count only those atoms which are mobile.

    Every Cd--Te bond in the system contributes an energy $\epsilon_c < 0$.
    The crystal structure is further stabilized by
    an effective intra-layer NN interaction $\epsilon_b$ of immobile
    particles of the same species.

    Effective interactions of the mobile particles at the surface
    are chosen to reproduce known properties of the system, e.g. surface reconstructions. 
    Consequently, mobile Cd and Te are treated on different footings.

    As in the planar models of Refs. \onlinecite{epl,flat} an infinite
    repulsion excludes  NN pairs of mobile Cd atoms in $y$-direction at the surface.
    The repulsion does not apply to pairs of one mobile and one immobile Cd, if 
    the latter has three or four Te partners. 
    An attractive interaction energy $\epsilon_x <0$ between mobile Cd-NN in $x$-direction and 
    a competing interaction $\epsilon_d<0$ between diagonal
    neighbors (NNN) favor the $(2\times1)$ or $c(2\times2)$ arrangement,
    respectively.  As demonstrated in Refs. \onlinecite{epl,flat}, the temperature dependence
    of the Cd reconstruction is reproduced qualitatively, if we set 
    $\left| \epsilon_x \right| \, \widetilde{\mbox{\small $<$}} \, 2\, \left| \epsilon_d \right| $.   

    NN pairs of mobile Te which do not form a dimer contribute an energy $\epsilon_m^{\rm Te} < 0$,
    whereas the energy $\epsilon_b^{\rm Td} < \epsilon_m^{\rm Te}$
    is assigned to a pair which forms a dimer.
    Note that dimerization introduces additional degrees of freedom to the model without 
    changing the topology of the lattice. 
    Pairwise interactions $\epsilon_x^{\rm Td}$ of neighboring  dimers in $x$-direction
    favor their arrangement in rows along $[110]$, corresponding to a $(2\times1)$
    surface symmetry.  For NN pairs of one mobile and one immobile Te
    we assume the same interaction $\epsilon_b$ as for pairs of incorporated atoms.

    In addition to Te particles at regular SOS lattice sites,
    we include the capture of Te atoms in a weakly bound state, denoted as Te$^*$ atoms
    in the following.  The energy contribution of each Te$^*$ atom is $\epsilon^*$
    where $\left|\epsilon^*\right| < 2 \left|\epsilon_c\right|$. This 
    accounts for the fact that the binding of a Te$^*$ is weaker than that of an isolated 
    mobile atom at a regular lattice site which has two Cd--Te bonds.

    Following an earlier hypothesis,\cite{tatarenko,cibert}
    these weakly bound Te$^*$ might be associated with atoms which occupy 
    Cd-sites temporarily and hence form a Te-trimer.\cite{ahrdiss} But, as already mentioned
    in section \ref{properties} the precise nature of the Te$^*$ is still unknown.
    Here, we abandon the assumption of Te-trimers and instead
    include Te$^*$ particles in an effective fashion.

    We introduce a reservoir of Te$^*$ with $\theta^* N$ particles,
    where $N$ is the number of atoms in one complete bulk layer of Te.  
    The total reservoir occupation $\theta^*$ is composed of two parts:
    $\theta^* = \theta^*_{loc} + \theta^*_{mob}$. The quantity 
    $\theta^*_{loc}$ accounts for the fraction of Te$^*$ which are localized on top of mobile Cd and thus 
    neutralize the infinite repulsion between Cd-NN in y-direction.\cite{ahrdiss}
    This effect will be discussed in more detail in the next 
    section. On the contrary, $\theta^*_{mob}$ corresponds to the remaining Te$^*$ atoms which are
    highly mobile and no assumption is made about their precise location and the nature of the
    binding. In general, during our simulations $\theta^*_{loc} \ll \theta^*_{mob}$ holds for most
    of the time.

    For the calculation of the Te coverage $\rho^{\rm Te}$ besides mobile Te atoms at regular lattice sites
    both localized and mobile Te$^*$ atoms are taken into account as well, i.e. $\rho^{\rm Te}$ 
    includes $\theta^*$ .
    According to the experimentally observed range of Te-coverages\cite{cibert,tatarenko2} 
    the total reservoir occupation thus is restricted to values $ 0 \leq \theta^* \leq 1/2$. 
    In our model we will exclude larger values by a simple
    cut-off: if $\theta^*$ has 
    reached its maximum value all processes leading to an increase of $\theta^*$
    are forbidden, i.e.  their rate is set to zero.

    We furthermore assume that an attractive pairwise interaction of Te$^*$ 
    stabilizes the reservoir. In a mean field fashion we represent the latter
    by an energy  contribution $N \, \epsilon_b^* \, {\theta^*}^2 $ with
    parameter $\epsilon_b^* <0$. 

   In summary, the total energy of a given configuration can be written as  
   \begin{eqnarray} \label{energy} 
    H  & = &  \epsilon_c \, n_c + \epsilon_b \left(n_b^{\rm Cd} +  n_b^{\rm Te} + n_h^{\rm Te}\right)
                 + \epsilon_x \, n_x^{\rm Cd} + \epsilon_d \, n_d^{\rm Cd} \nonumber \\ 
       &+ &   \epsilon_m^{\rm Te} \, n_m^{\rm Te} + 
       \epsilon_b^{\rm Td} \, n_b^{\rm Td} + \epsilon_x^{\rm Td} \, n_x^{\rm Td} \nonumber \\
       &+ & \epsilon^* \theta^* N + \epsilon_b^* {\theta^*}^2 N 
  \end{eqnarray}
  which depends on the number of Cd-Te bonds ($n_c$),
  bulk NN pairs of Cd ($n_b^{\rm Cd}$) and Te ($n_b^{\rm Te}$), NN pairs of one mobile and
  one immobile Te ($n_h^{\rm Te}$),
  NN and NNN pairs of mobile Cd  at the surface ($n_x^{\rm Cd}$ and $n_d^{\rm Cd}$),
  and NN pairs of mobile Te not forming a dimer ($n_m^{\rm Te}$). The number of  Te-dimers is denoted
  as $n_b^{\rm Td}$, whereas $n_x^{\rm Td}$ counts pairs of neighboring 
  Te-dimers along the $x$-direction.  The last line in Eq.\ (\ref{energy})
  accounts for the contribution of the 
  Te$^*$ reservoir which is occupied by $\theta^* N$ atoms. 

  Throughout this paper we refer to the following parameter set:
  \begin{eqnarray} \label{parameters1}
  \epsilon_d = -1, \,\, \epsilon_x = -1.95, \,\, \epsilon_c = -6.5, \,\, \epsilon^* = -5, \,\, \epsilon_b = -0.8, \nonumber\\
  \epsilon_b^* = -8, \,\, \epsilon_m^{\rm Te} = -0.6, \,\, \epsilon_b^{\rm Td} = -1, \,\, \epsilon_x^{\rm Td} = -0.4 \qquad
  \end{eqnarray} 
  Note that $\left| \epsilon_d \right|$ only fixes the energy scale in the following.
  In addition 
  we set $k_B=\left|\epsilon_d\right|=1$ formally and, hence,  measure the temperature $T$ 
  on the same scale.  Later we will comment on a 
  potential quantitative comparison with experiments.   

%
%

 \subsection{Dynamics and growth}

  As we are interested in the non-equilibrium dynamics of the system under 
  sublimation or growth conditions, specifying the energy of a given configuration
  is not sufficient.  
  In fact, only the energy contribution of a few active layers which contain
  mobile particles will be relevant in the following. 
  The kinetics of the system is governed  by 
  the energy barriers which have to be overcome
  in transitions between the different configurations.   

  Unfortunately, the available experimental data and first principles results
  are not sufficient for a systematic fit of our model parameters. 
  In particular, reliable evaluations or estimates of, e.g.,  diffusion
  barriers  are not available,  apart from very few exceptions.  
  Therefore, we do not aim at a precise quantitative description of
  the CdTe(001) surface at this stage.  

  Nevertheless, essential features can be deduced  from experimental data and
  physical insight. Important qualitative features of the
  model turn out to be quite robust against variations of the parameters,
  as long as they comply with some essential conditions. 
  In our simulations we will use a parameter set which is mainly
  based on our  previous investigations of related models and their
  comparison with experimental findings. 

  In our previous studies of simpler models, appropriate choices of the parameter 
  set enabled us to reproduce qualitatively the observed temperature
  dependent structure of Cd-terminated surfaces in step-flow sublimation. Furthermore, a 
  semi-quantitative match of the re-ordering temperature, macroscopic sublimation rates etc. 
  has been achieved, see Refs. \onlinecite{aachen,ahrdiss}.

  The microscopic processes considered in our Kinetic Monte Carlo (KMC) dynamics are
  deposition of Cd and Te adatoms, diffusion and desorption of mobile atoms,
  transitions of Te atoms between regular lattice sites and the reservoir, 
  and desorption of reservoir atoms.

  Deposition of Cd and Te atoms occurs with fluxes $F^{\rm Cd}$ and $F^{\rm Te}$, 
  respectively.
  First, a lattice site is chosen randomly. Then, 
  a search is performed within an 
  incorporation radius $r_{inc}=2$ for the site with the lowest height where adsorption is possible.\cite{ahrdiss}
  If this is, for a Cd atom, a site with a neighboring mobile Cd atom in y-direction, adsorption is only
  possible if there is either a localized Te$^*$ on top of the Cd neighbor, already,
  or a mobile Te$^*$ attaches
  to the impinging Cd. This occurs with probability $\theta^*_{mob}$.
  For a Te atom proper adsorption sites include, besides regular Te sites, mobile Cd atoms with
  no localized Te$^*$ attached as well as empty Cd sites. In the latter two cases, the Te then becomes a 
  mobile reservoir Te$^*$ with probability $1 - 2 \theta^*$, whereas with probability
  $2 \theta^*$ deposition is rejected and the particle evaporates. As a consequence, the
  reservoir occupation is limited to values $\theta^* \leq 1/2$.

\begin{table}
\caption{\label{tabelle}
Thermally activated processes in the simulation and corresponding values of attempt frequencies 
    $\nu_i$, prefactors $\alpha_i$ and activation energies $E_i$ used for calculating
    the Arrhenius rates (Eq.\ (\ref{arrhenius})) as described in the text.
    Parameters used in our simulations are $\nu_o=10^{12} s^{-1}$, $\nu^*=10^8 s^{-1}$, $B_o=7$, $B^*=2$.
    The energy change $\Delta H_i$ for a specific event is given by Eq.\ (\ref{energy}).
}
\begin{ruledtabular}
\begin{tabular}{lccc}
 process $i$ & $\nu_i$ & $\alpha_i$ & $E_i$\\
\hline
Cd or Te diffusion & $\nu_o$ & $1$ & $\max \left\{ B_o , B_o + \Delta H_i \right\}$ \\ \hline
    Cd diffusion (Te$^*$ involved) & $\nu_o$ & $\theta^*_{mob}$ & $\max \left\{ B_o , B_o + \Delta H_i \right\}$ \\ \hline
    Cd or Te desorption & $\nu_o$ & $1$ & $\Delta H_i$ \\ \hline
    Te $\rightarrow$ Te$^*_{mob}$ & $\nu_o$ & $1 - 2 \theta^*$ & $\max \left\{ B_o , B^* + \Delta H_i \right\}$ \\ \hline
    Te $\rightarrow$ Te$^*_{loc}$ & $\nu_o$ & $1 - 2 \theta^*$ & $\max \left\{ B_o , B^* + \Delta H_i \right\}$ \\ \hline
    Te$^*_{mob}$ $\rightarrow$ Te & $\nu^*$ & $\theta^*_{mob}$ & $\max \left\{ B^* , B_o + \Delta H_i \right\}$ \\ \hline
    Te$^*_{loc}$ $\rightarrow$ Te & $\nu^*$ & $1$ & $\max \left\{ B^* , B_o + \Delta H_i \right\}$ \\ \hline
    Te$^*_{mob}$ desorption & $\nu^*$ & $\theta^*_{mob}$ & $\Delta H_i$ \\
\end{tabular}
\end{ruledtabular}
\end{table}

  The rates for all thermally activated processes are of the Arrhenius form
  \begin{equation} \label{arrhenius} 
    r_i \, = \, \nu_i \,\, \alpha_i \,\, \exp \left( - E_i / T \right),
  \end{equation} 
  where $E_i$ and $\nu_i$ denote the activation energy and attempt frequency for process $i$, respectively.
  The prefactor $\alpha_i$ accounts for rates which depend explicitly on the reservoir occupation 
  (cf. Table \ref{tabelle}).
  To calculate the activation energy $E_i$ we assume the energy of the transition state 
  $H_{t,i}$ between
  the starting ($s$) and the final ($f$)  
  configuration of the system to be equal to the greater of the energies $H_{s,i}$
  and $H_{f,i}$ of both configurations plus an additional barrier:\cite{newman99}
  $H_{t,i} = \max \left\{ H_{s,i} + B_{s,i}, H_{f,i} + B_{f,i} \right\}$.
  The values of the barrier heights $B_{s,i}$ and $B_{f,i}$ in general are not the same but depend on the
  type of process $i$. For
  the activation energies $E_i = H_{t,i} - H_{s,i}$ one obtains
  \begin{equation} \label{activation}
    E_i = \max \left\{ B_{s,i}, B_{f,i} + \Delta H_i \right\}
  \end{equation}
  for diffusion of mobile atoms and transitions of Te atoms between regular lattice sites 
  and reservoir states
  with the energy change $\Delta H_i = H_{f,i} - H_{s,i}$ given by Eq.\ (\ref{energy}).
  Diffusion includes moves to available NN or NNN sites in the same layer, as well as hops across
  terrace edges, cf. Figure \ref{sosbild}.
  For all diffusion steps we set $B_{s,i} = B_{f,i} = B_o$ in Eq.\ (\ref{activation}) with barrier $B_o = 7$, 
  similar to Refs. \onlinecite{aachen,ahrdiss}. On the contrary, the assumption that Te$^*$ atoms in the reservoir are 
  less strongly bound implies $B_{s,i} = B_o$ and $B_{f,i} = B^* < B_o$ for the transition of a Te atom 
  into the reservoir and, conversely, $B_{s,i} = B^*$ and $B_{f,i} = B_o$ for the opposite process 
  (cf. Table \ref{tabelle}). 

  Desorption of atoms from regular lattice sites or mobile Te$^*$ requires an activation energy 
  $E_i = \Delta H_i$.
  We consider only desorption of single atoms, neglecting the fact that Te probably evaporates from
  the surface in the form of Te$_2$ molecules, in reality.

  For all diffusion processes and the desorption of mobile atoms at regular surface sites we assume an
  attempt frequency $\nu_i = \nu_o = 10^{12} s^{-1}$, whereas a lower frequency $\nu_i = \nu^*$
  is used for the transition of Te$^*$ atoms to regular Te states and the 
  evaporation of mobile Te$^*$.

  This choice is based on the reasonable
  assumption that Te$^*$ atoms reside in potential energy minima 
  which are much shallower than
  those for Te atoms at regular surface sites.\cite{zangwill} According to transition state 
  theory the attempt frequency $\nu^*$ which corresponds to the harmonic oscillation frequency
  at the local energy minimum will become smaller.\cite{tst} 
  A similar argument was used
  by Itoh \cite{itoh01} in order to motivate an attempt frequency of 
  weakly bound arsenic as low as 
  $10^8 s^{-1}$. In our simulations, we will also set $\nu^* = 10^8 s^{-1}$.
  This choice is further motivated by comparison with experimental observations as 
  discussed in the following section.

  The formation or breaking of dimers from NN pairs of mobile Te atoms is
  taken into account
  implicitly. A diffusion hop of a Cd atom onto a Te dimer,
  for instance, requires the breaking of the dimer which has to be considered in the
  energy balance $\Delta H_i$ of the process. The deposition of a Cd atom will always break
  all underlying Te dimers involved.

  On the other hand, dimer formation is assumed to occur instantaneously, whenever it 
  is possible  due to the resulting energy gain.
  A similar simplification was used in Ref. \onlinecite{barnett} 
  for the description of Si-dimerization on Si(001) surfaces. 
  Hence, if any event results in two un-dimerized Te atoms
  which are in the correct configuration for forming a dimer pair, a dimer bond will be created
  immediately. For diffusion and desorption events, dimer formation will be accounted for in $\Delta H_i$.

  The existence of a weakly bound state for Te atoms is essential to
  facilitate growth in our model. Consider a Cd-terminated surface with $c(2\times2)$ reconstruction 
  and maximum
  coverage $\rho^{\rm Cd} = 1/2$: clearly, deposition of Te at a regular lattice site
  would violate the SOS condition. On the other hand, the incorporation of additional Cd
  is impossible due to the infinite repulsion of NN neighbors in $y$-direction.

  It is plausible to assume, however, that an additional Te bound to a Cd
  would neutralize the repulsion as it provides three negatively 
  charged dangling bonds.
  In our model, mobile Cd can hop to empty sites neighboring another mobile
  Cd in $y$-direction, provided there is either already a Te$^*$ localized on top of the neighboring Cd, 
  or a mobile Te$^*$ attaches to the diffusing particle. This Te$^*$ then would become localized itself. 
  The latter scenario is assumed to happen with a
  probability $\theta^*_{mob}$ which is reflected in the prefactor $\alpha_i$ of the corresponding rate
  (cf. Table \ref{tabelle}).

  Thus, we assume that mobile Te$^*$ which are not attached to Cd atoms 
  are available instantaneously everywhere at the surface.
  Note that after the diffusion step
  the reservoir occupation $\theta^* = \theta^*_{loc} + \theta^*_{mob}$ remains the same 
  but the ratio $\theta^*_{loc} / \theta^*_{mob}$ has increased. If at any point of the
  simulation a localized Te$^*$ is no longer needed because for example the underlying Cd atom
  has evaporated, it will become a mobile Te$^*$ again.
  Through the mediation of Te$^*$ atoms, Cd atoms at the surface can coalesce into islands
  in spite of the NN repulsion, provided the reservoir of Te$^*$ is not empty.
  
  In our simulations we consider also transitions of Te between regular lattice sites and reservoir 
  states as well as the desorption of mobile Te$^*$ into the vacuum. In general, the rates of 
  these processes 
  depend on the reservoir occupation, i.e. $\alpha_i \neq 1$ (cf. Table \ref{tabelle}). 
  For instance a mobile Te at the surface may become
  a mobile Te$^*$ provided the reservoir is not fully occupied. This happens with a rate
  according to Eq.\ (\ref{arrhenius}) with  $\alpha_i = \left( 1 - 2 \theta^* \right)$
  which is zero if the reservoir density has reached its maximum
  value $1/2$. Conversely, the transition of a mobile Te$^*$ to a regular Te state is proportional 
  to the density 
  $\theta^*_{mob}$ of mobile Te$^*$ atoms. Localized Te$^*$ atoms which are on top of Cd atoms in order
  to neutralize the infinite repulsion may become bound at one of the two neighboring regular Te sites, 
  if available, thus reducing the density $\theta^*_{loc}$.

 
  The results presented in the following sections were obtained in simulations
  with energy parameters according to Eq.\ (\ref{parameters1}) and the 
  following barriers and frequencies: 
  \begin{equation} \label{parameters2}
  \nu_o=10^{12} s^{-1}, \,\,\, \nu^* = 10^{8} s^{-1}, \,\, B_o = 7, \,\,\,  B^* = 2 
  \end{equation}

%
%

 \section{Simulations and results} \label{results} 

 \subsection{The reservoir of excess Te$^*$} \label{reservoir}

  Before presenting results of the ALE  simulations we discuss
  the properties of the Te$^*$ reservoir in greater detail.
  In our model, the  occupation $\theta^*$ 
  results from the complex interplay of deposition, evaporation and transitions
  between Te$^*$ and Te states. 
  However, the basic temperature dependence of $\theta^*$ can be understood
  from considering a flat surface in the presence of a steady flux of Te atoms. 

  After the Te coverage has reached the value $\rho^{\rm Te}=1$, only deposition
  directly into the reservoir is possible. If we neglect the desorption of regular Te 
  from the surface and transitions of the type Te $\to$ Te$^*$, 
  the temporal evolution of $\theta^*$ is governed by
  the balance of direct deposition and evaporation of Te$^*$. The rate of the latter
  is a function of $\theta^*$ itself and one obtains the 
  following simplifying ODE: 
  \begin{equation} \label{thetadgl}
     \frac{d \, \theta^*}{d \, t} \, \approx \,  F^{\rm Te} \, \Theta \left[ 1/2 - \theta^* \right] \, - \theta^* \, \nu^* \,
       \, \exp \left[ {- \frac{1}{T} \left( \left| \epsilon^* \right| + \theta^* \left| \epsilon_b^* \right| \right)} \right] 
  \end{equation}
  Whereas the Te flux increases $\theta^*$, desorption of Te$^*$ occurs with
  a frequency proportional to $\theta^*$ multiplied with the Arrhenius rate  
  for single particle evaporation from the reservoir.   
  Clearly, the filling of the reservoir must saturate at $\theta^* \approx 1/2$.
  However, the precise nature of the saturation process is not essential for the following arguments and
  we choose a simple cut-off represented by the Heaviside-function $\Theta[x]$. 

  Given $\epsilon^*, \epsilon_b^*$ and $F^{\rm Te}$, 
  the right hand side in Eq.\ (\ref{thetadgl}) is positive for all $\theta^* \in \left[ 0,1/2 \right]$ 
  if the temperature is low enough. Hence, 
  the reservoir will be filled in this case. 
  Above a characteristic temperature $T_{res}$ the r.h.s.\ displays two zeros at
  fairly small values of $\theta^*$.   As a consequence, the  differential
  equation has an attractive fixed point with a stationary value $\theta^* \approx 0$
  for $T>T_{res}$. 

\begin{figure}[t]
    \begin{center}
    \includegraphics{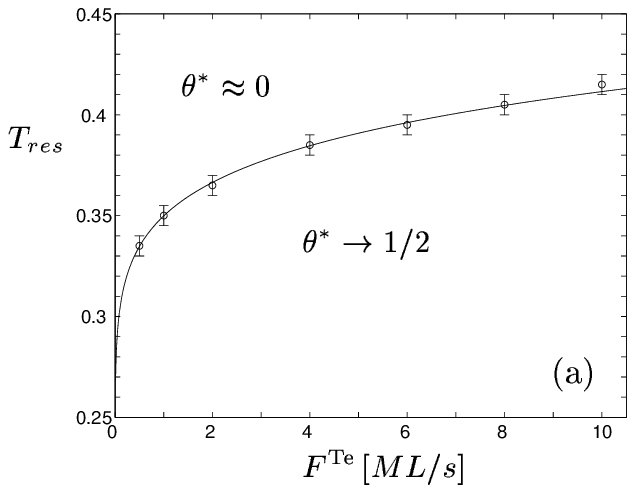} 
    \includegraphics{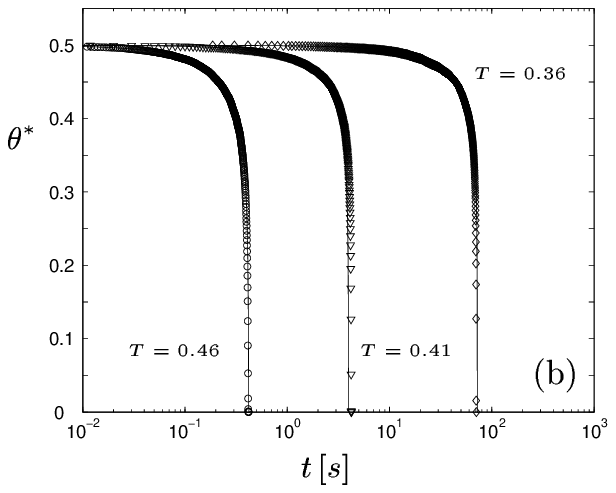}
    \caption{ \label{tresvsf}  
      Analysis of the reservoir dynamics 
    for  $\epsilon^* = -5, \epsilon_b^* =-8$.\protect\newline
     (a) The temperature $T_{res}$ above
    which an initially empty reservoir of Te$^*$ cannot be filled. The curve
    shows the result as obtained from Eq.\ (\ref{thetadgl}).
    Additionally, symbols indicate the temperature where the growth rate drops from $\approx 1ML$
    to $\approx 0.5 ML$ CdTe per ALE cycle in the KMC simulations (cf. section \ref{ale}).\protect\newline
    (b) Sublimation of Te$^*$ from an initially filled reservoir  
    for three different temperatures. Solid lines represent the numerical
    integration of Eq.\ (\ref{thetadgl}) with $F^{\rm Te} = 0$, 
    symbols correspond to simulations of the situation in the full model.  \protect\newline
    }
    \end{center}
    \end{figure}

  Figure \ref{tresvsf} (left) shows $T_{res}$ as function of $F^{\rm Te}$ for the
  choice of parameters $\epsilon^* = -5$ and $\epsilon_b^* = -8$. There is a rapid increase
  of $T_{res}$ for very low values of $F^{\rm Te}$ which becomes weaker and weaker for higher
  fluxes. With $F=5ML/s$ we obtain $T_{res} \approx 0.39$,
  as an example,  and at $T = 0.40$ the stationary occupation
  is as low as $\theta^* \approx 0.02$. 
  Hence, starting from $\theta^* =0$, the reservoir cannot be filled
  significantly for $T> T_{res}$.  
  The behavior predicted from Eq.\ (\ref{thetadgl}) is in excellent agreement
  with simulations of the situation in the full model. 
  We would like to mention that in Ref. \onlinecite{faschinger} a simple evaporation model 
  of ALE growth was proposed
  which produces a flux-dependence of the characteristic temperature 
  which is similar to the one in Figure \ref{tresvsf} (left).
 
    \begin{figure*}[t]
      \begin{center}
        \includegraphics[width=0.99\textwidth]{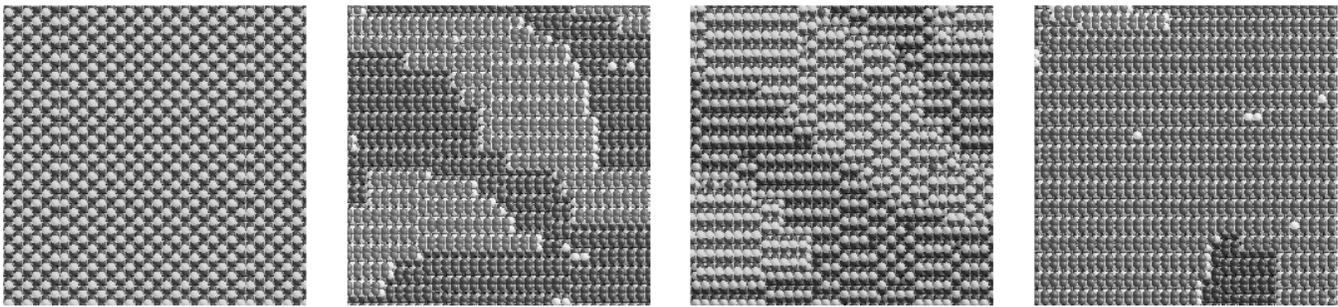} 
        \caption{ \label{seq_dg44} ALE simulations with $T=0.44$, snapshots of
          the surface at (left to right) $t=1s, \, 2s, \, 3s, \, 4s$ corresponding to the end
          of stages I,II,III,IV in the high temperature growth scenario as described in the text.
          Dark / light grey spheres correspond to Te / Cd atoms, respectively. Darker portions of
          the surface correspond to lower height above the substrate. Te-dimers are represented
          by smaller lateral distance of NN atoms.}
      \end{center}
    \end{figure*}

    The simplified Eq.\ (\ref{thetadgl}) is also useful in investigating the initial
    sublimation of Te$^*$ from a surface with maximum coverage $\rho^{\rm Te} = 3/2$, 
    i.e.\  $\theta^* = 1/2$. 
    Figure \ref{tresvsf} (right)
    displays the evolution of $\theta^*$  for $\epsilon^* = -5, \epsilon_b^* =-8$ and
    three  different temperatures 
    as obtained from the numerical integration of (\ref{thetadgl}) with $F^{\rm Te} = 0$.
    The results from simulations of the corresponding situation in the full kinetic
    model are again in excellent agreement with the simplified description, which neglects
    evaporation of regular Te and other processes. 
    
    One obtains a well-defined characteristic time $\tau^*$ for the decrease of $\theta^*$ from
    $0.5$ to, say, $\theta_\tau^* = 0.05$.  The precise value of $\theta^*_\tau$ is of 
    little relevance  as $\theta^*(t)$ decreases to zero very rapidly, 
    cf. Figure \ref{tresvsf} (right panel). 
    The temperature dependence of $\tau^*$  is very well described by an
    Arrhenius law of the form $\tau^* \, = \, \tau_o \, e^{E_a/T}$  with  
    macroscopic activation energy $E_a \approx 8.5$ and prefactor $\tau_o \approx 3.8\times10^{-9} s$. 
    In Refs. \onlinecite{tatarenko,cibert} experimental data is discussed for the
    initial sublimation of Te starting from coverage $3/2$. There, the macroscopic activation
    energy is reported to be $E_a \approx 0.96 \, eV$,
    providing a potential reference point for attempting a quantitative fit of the model. 

    Note that the prefactor $\tau_o \sim 10^{-9} s$  obtained from Fig. \ref{tresvsf} (right)
    is not too different from the time constant $1/\nu^* = 10^{-8} s$
    assumed  for Te$^*$ desorption.
    This should be expected, as the latter is the limiting microscopic process of the scenario. 
    In a naive attempt to extract the prefactor from an analogous Arrhenius plot of experimental data
    \cite{tatarenko} we obtain  a value of about the same order, i.e. $\tau_o \sim 10^{-9} s$. 
    On the contrary, the value 
    from data for Cd desorption \cite{tatarenko} corresponds to frequencies which
    are larger  by  orders of magnitude. Of course, a reliable determination of prefactors from 
    experimental data for relatively small temperature ranges is questionable if not 
    impossible (and was not attempted by the authors of Ref. \onlinecite{tatarenko}). 
    Nevertheless one might interpret the very rough comparison as another
    qualitative justification for choosing $\nu^*$ much smaller than other 
    frequencies in the model. 

%
%

 \subsection{ALE growth} \label{ale}

 In order to gain a qualitative understanding of the ALE growth scenario
 we perform simulations with the parameter set specified in
  Eqs.\  (\ref{parameters1},\ref{parameters2}). All simulation
 results presented here were obtained from systems with 
 $64\times64$ Cd and Te atoms in a bulk layer, averaged
 over $5$ independent runs. 

 We model  a situation in which Cd and Te are deposited in alternate pulses of 
 length  $0.9s$, each with a constant flux $F^{\rm Cd} = F^{\rm Te} = 5 ML/s$. 
 The pulses are separated by a dead time of $0.1 s$,
 hence, the duration of one complete ALE cycle is $2s$ in total. 
 Note that 
 the characteristic time $\tau^*$ for reservoir sublimation
 is large compared with the dead time interval for all relevant temperatures. 
 In the results discussed here, the initial surface was perfectly flat 
 and Te-terminated 
 with coverage $\rho^{\rm Te} = 1$, $\theta^*=0$. Deposition always started with
 a Cd pulse.

\begin{figure*}[t]
  \begin{center}
    \includegraphics[width=0.99\textwidth]{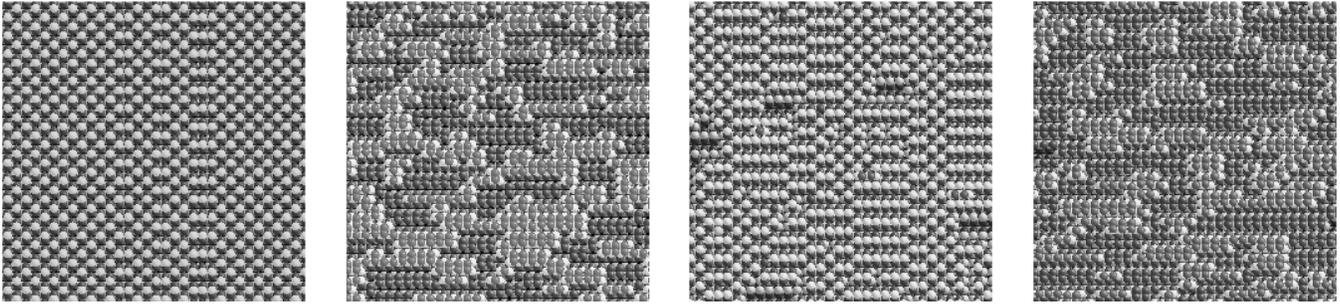} 
    \caption{ \label{seq_dg36} ALE simulations with $T = 0.36$, snapshots of the surface at (left
      to right) $t=1s, \, 2s, \, 3s, \, 4s$, the two last corresponding to the end of stages A,B in the low
      temperature growth scenario as described in the text.}
  \end{center}
\end{figure*}

 Illustrating surface snapshots of growing surfaces are shown
 in Figures \ref{seq_dg44} and \ref{seq_dg36} for the two different growth regimes
 discussed in the following. Additional illustrations and mpeg-movies are available from
 our web pages.\cite{web}

\begin{figure}[t]
    \begin{center}
    \includegraphics{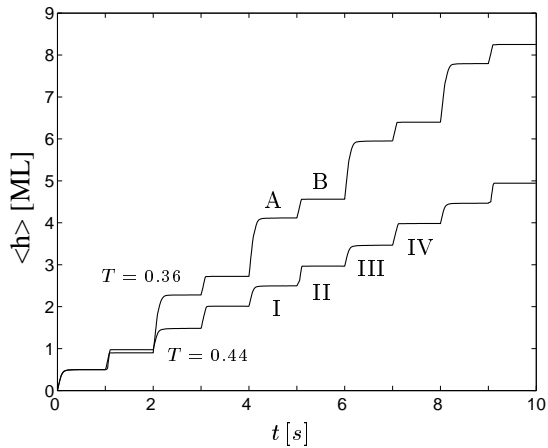} 
    \caption{ \label{hvst}
     Mean height of the film vs. time in the simulations of ALE for two different substrate  
     temperatures. The pulse duration is $0.9s$ with $F^{\rm Te} = F^{\rm Cd} = 5 ML/s$
     followed by a dead time of $0.1 s$. 
     $\left\langle h \right\rangle$ takes into account only particles at
     regular lattice sites, excluding the reservoir of Te$^*$. 
     The upper curve evolves according to the low temperature scenario ($T=0.36$) with steps
     A and  B whereas the lower corresponds to  $T=0.44$ 
     with the double cycle sequence I,II,III,IV  as described in the text.
     The scenarios essentially coincide for the first $2 s$ because in both cases
     the reservoir is empty before the first Cd-pulse is deposited.}
    \end{center}
    \end{figure}

 At fairly high temperatures, growth proceeds according to 
 the following scenario  in our simulations:
 \begin{itemize}
 \item[I)] The first Cd pulse adds half a layer of Cd to the system. 
       The infinite repulsion
       of NN in $y$-direction prevents coverages $\rho^{\rm Cd} > 1/2$. 
       The predominant arrangement of Cd atoms is in a local $c(2\times 2)$ 
       ordering, i.e.\ a checkerboard pattern. The contribution of the $(2\times1)$
       row structure as a thermal excitation increases with higher T.
 \item[II)]  After the first Te pulse and the dead time, half a layer of Te has been
       incorporated.  Diffusion of particles  
       leads to the formation of islands upon 
       the Te-terminated surface.  Note that the existence of the Te$^*$-reservoir state
       is crucial for this process, as discussed above.
       Surface Te atoms form dimers, preferentially, and
       these arrange in rows along the $x$-direction, yielding a predominant
       $(2\times1)$ symmetry of the surface.  
       Although incorporation of Te occurs to a considerable degree through the reservoir state Te$^*$,
       at the end of the dead time one finds $\theta^* \approx 0$, as discussed above.  
 \item[III)] The second Cd pulse places mobile atoms on top of existing islands 
       and in between.  The pre-dominant local ordering is the $(2\times1)$ pattern,
       i.e.\ rows of Cd alternating with empty rows. This is due to the influence of the
       island edges on the reconstruction, see Refs. \onlinecite{aachen,surflett,ahrdiss} for a discussion
       of an analogous effect in layer by layer sublimation.
       At the end of the dead time,
       the islands from stage II are still present but now they are Cd-terminated.  
 \item[IV)]
       During the second Te pulse, atoms are deposited onto
       the Cd-terminated islands and in between. The particles rearrange by means of
       diffusion, leveling off the islands and forming a flat Te-terminated
       surface.  
       At the end of the dead time, the initial state of the system is restored
       with one complete layer of CdTe added to the system. 
       In the simulations the filling of gaps is  incomplete
       to a degree which depends on the temperature and the length of the 
       dead time interval. 
 \end{itemize}
Hence, at temperatures $T> T_{res}$, our model reproduces ALE growth in the same 
double cycle sequence which was hypothesized and discussed for CdTe and ZnTe 
in the literature, see Refs. \onlinecite{cibert,tatarenko2} and references therein. 
Figure \ref{hvst} shows the evolution of the film height with time in 
the ALE process at $T = 0.44$, which is well above the characteristic
$T_{res} \approx 0.39$ obtained for $F^{\rm Te} = 5 ML/s$. 

The picture changes qualitatively for temperatures $T < T_{res}$.
Again, a flat Te-terminated surface with
coverage $\rho^{\rm Te} = 1$, $\theta^* = 0$ was prepared, 
and the deposition started with a Cd pulse.
The surface structures obtained after the first Cd and Te pulse are very much alike
the ones of stages I and II of the high temperature scenario, since initially
the Te$^*$ reservoir was empty. Note however, that due to a decreased
mobility of the atoms the islands on the Te terminated surface are less compact.
Since $T < T_{res}$, at the end of the Te pulse the surface is not only Te terminated 
but also the Te$^*$ reservoir is filled, i.e. $\theta^* \approx 1/2$.

Thus, from this point on the ALE growth scenario proceeds along the following lines
in our model:

 \begin{itemize}
 \item[A)]   The  Cd pulse places atoms onto the islands and in between which is analogous
             to stage III of the high temperature scenario. But now Cd particles at the surface 
             can coalesce, because Te$^*$ from the reservoir are available
             to neutralize the NN repulsion of mobile Cd in $y$-direction. Therefore
             the gaps between the islands are leveled out by half a monolayer of Cd atoms
             and the reservoir which provides also approx.\  $1/2$ monolayer of Te.
             The surface is then covered by another half layer of Cd atoms arranging
             preferentially in a $c(2\times2)$ vacancy structure. Thus, the adsorption of one complete 
             monolayer of Cd and a half monolayer of Te which came from the reservoir leads to
             an increase of the surface height by $3/4$ of a complete CdTe layer. At the end
             of step A) the reservoir is empty.
 \item[B)]   The following Te pulse leads to a Te terminated surface with islands, which is identical 
             to the one described above. Again, due to low temperature and short dead time, also 
             the Te$^*$ reservoir is filled. Hence, another half monolayer of Te is adsorbed 
             and in total one complete layer of CdTe added in only one ALE cycle. At the end
             of step B) the reservoir has reached its maximum occupation.
 \end{itemize}

 As an example, Figure \ref{hvst} shows the evolution of the mean surface
 height $\left< h \right>$ with time at $T = 0.36$, i.e. below $T_{res}$ for $F^{\rm Te} = 5 ML/s$. 
 Note that for the first Cd and Te pulse, the curve is essentially identical
 with that of the high $T$ regime, because in both simulations the reservoir
 was empty, initially.
 Hence the upper curve of Figure \ref{hvst}  corresponds to the sequence of  stages A/B only
 for times $t > 2 s$. 
 The  mean height as displayed in Figure \ref{hvst}  
 increases by $3/4$ of a complete CdTe layer during the Cd pulse A)  because $\left< h \right>$ takes
 into account only particles at SOS lattice sites, disregarding the Te$^*$ reservoir. 

\begin{figure}[t]
    \begin{center}
      \includegraphics{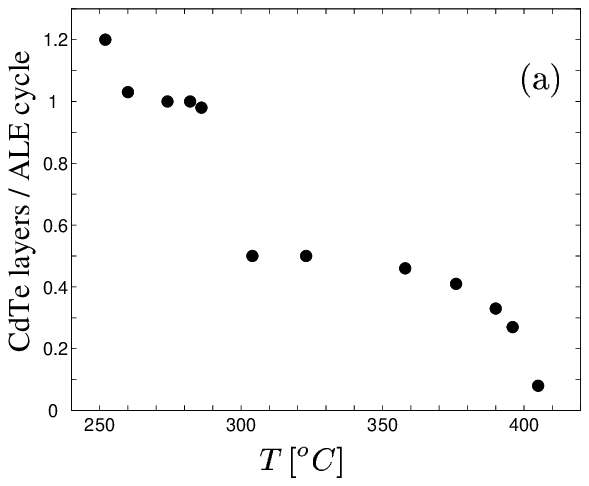} 
    \includegraphics{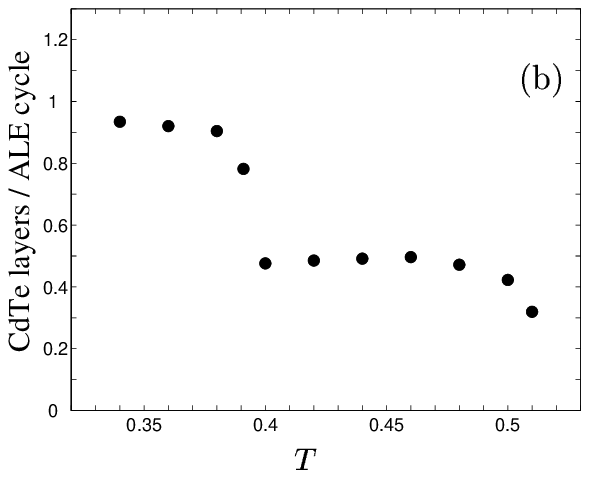} 
    \caption{\label{rates} 
     Growth rates in layers of CdTe per ALE cycle {\sl vs.} temperature $T$.\protect\newline
     (a) Experimentally observed growth rates of CdTe(001) 
     as published 
     by Faschinger and Sitter, plot generated after
      Fig.\ 3 in Ref. \onlinecite{faschinger}. 
     The pulse duration was $0.8s$ with a total deposition of 1.5 layers of
     each element, followed by dead time intervals of length $0.2s$.  
     The most prominent feature is the sudden decrease of the growth rate
     at $T\approx 290^o C$.  One additional data point has been omitted 
     here (growth rate 1.4 layers per cycle at 230$^oC$).    
     \protect\newline
     (b) Average growth rate in our simulations with parameters as in
       Fig.\ \ref{hvst} for different temperatures. $T$ is measured 
       in dimensionless units of $\left| \epsilon_d \right| = 1$. 
       The growth rate drops at $T\approx T_{res} \approx
       0.39$ for $F^{\rm Te} = 5 ML/s$.  }
    \end{center}
    \end{figure}

 From simulations for different temperatures we have obtained the 
 average growth rate per ALE cycle. 
 Figure \ref{rates}(b) shows that, with increasing $T$, 
 the growth rate drops  very rapidly  in the vicinity
 of the characteristic $T_{res}$  as obtained from the considerations of
 section \ref{reservoir}. 
 The comparison with Figure \ref{rates}(a) 
 shows qualitative agreement with experimental
 results for an analogous growth scenario.\cite{faschinger}
 Very similar  data has been published by Hartmann et {\sl al.}\cite{hartmann}
 and is also discussed in Ref. \onlinecite{cibert}.  
 Note that in both experiments growth rates above 1 layer
 of CdTe per ALE cycle were observed at very low temperatures. 
 This is presumably due to the formation of Cd crystallites at the surface and cannot
 be reproduced in our model.\cite{tatarenko2}

 The temperature driven break-down of the growth rate is due to the fact that
 the reservoir cannot be filled significantly for $T > T_{res}$. Simulation results
 for various values of the flux strongly support this assumption as can be seen
 from Figure \ref{tresvsf} (left): the symbols which indicate the temperature where the
 growth rate drops in the simulations 
 lie almost exactly on the curve that marks $T_{res}$.

 In principle, the re-evaporation of Te$^*$ during the dead time interval 
 could be an alternative cause for the reduced growth rate. 
 Note, however, that even for a flux as 
 high as $F^{\rm Te} = 10 ML/s$ with  $T_{res} \approx 0.41$,
 one finds that the characteristic time for reservoir sublimation is $\tau^* \approx 4 s$,
 cf.\ Figure \ref{tresvsf} (right panel).
 Hence, it appears safe to say that during dead time intervals shorter than, say, $1 s$,  
 $\theta^*$ will remain close to the maximum value $1/2$, provided all other 
 conditions allow for the filling of the reservoir during a Te pulse. 

 As a consequence, we expect that the characteristic 
 value of $T$ which marks the drop
 of the growth rate in Figure \ref{rates} should be essentially 
 independent of the pulse duration or dead time interval 
 within a wide range of reasonable values. \\

 Comparing the experimental results by Hartmann et {\sl al.}\cite{hartmann} and 
 Faschinger and Sitter\cite{faschinger} 
 for the ALE growth rate one finds that for the latter the temperature where the growth
 rate drops to approx.\ $1/2$ layers of CdTe per cycle 
 is shifted to higher temperatures by about $40 K$. Within  the
 framework of our model this can be  explained if one takes into account the different
 experimental conditions. Whereas Hartmann et {\sl al.} combined particle fluxes of about $0.5 ML/s$ 
 with $8 s$ pulse time and $1 s$ dead times, the flux used by Faschinger and Sitter was $2 ML/s$
 with a pulse duration of $0.8 s$ and a dead time of $0.2 s$. As discussed above, the difference
 of the dead times can be neglected, but for the given fluxes one reads off from 
 Fig.\ \ref{tresvsf} (left) a significant increase of $T_{res}$ by about $9 \%$. As demonstrated,
 $T_{res}$ practically coincides with the temperature where the growth rate drops which consequently
 is expected to increase by the same amount. The difference of the absolute 
 temperatures observed in Refs. \onlinecite{hartmann} and \onlinecite{faschinger} is about $8 \%$ 
 (of the lower one) which 
 is indeed comparable with our results.
 
 For very high $T$, 
 sublimation of the crystal dominates over the incoming flux
 and growth becomes impossible. The ALE growth rate vanishes at a
 temperature which is essentially independent of the properties of 
 Te$^*$ atoms.

%
%

  \section{Summary and Outlook} \label{discussion}  

  In summary, we have presented a detailed lattice gas model of
  II-VI semiconductor compounds. In contrast to many simpler growth
  models, it takes into account the correct zinc-blende lattice as well
  as essential, material specific effects which have been discussed
  in terms of the CdTe(001) surface, as an example.
  The model reproduces, for instance, 
  the reconstruction of Cd-terminated surfaces by means of vacancy structures 
  and its dependency on temperature and flux. This includes 
  the temperature driven transition from the $c(2\times2)$ reconstructed surface to
  a disordered state with local $(2\times1)$ ordering.  This re-arrangement  of
  the surface is observed under step flow sublimation, as demonstrated earlier in  
  Refs. \onlinecite{aachen,surflett}, and a semi-quantitative agreement with experimental
  data can be achieved. 

  Here, in addition, the dimerization of Te at the surface
  is introduced to the model. Thus, the $(2\times1)$ reconstruction of
  Te-terminated surfaces is reproduced. 
  We demonstrate furthermore that the presence of 
  weakly bound Te$^*$ atoms is necessary to facilitate growth by, e.g., 
  atomic layer epitaxy. Although the precise nature of such a weakly bound
  Te$^*$ state at the surface is unknown, its existence is clearly evident from 
  various experimental observations.\cite{cibert,tatarenko2}

  By means of Kinetic Monte Carlo simulations of our model we show that
  the reservoir of Te$^*$ also explains the experimentally observed 
  temperature dependence of the ALE growth rates.  Only below a characteristic
  temperature a significant amount of Te$^*$ is present and facilitates 
  growth rates of  approximately $1$ layer of CdTe per ALE cycle. As in the corresponding
  experiments, we observe a sudden transition from the self-limited growth rate of about $1/2$ 
  layer per ALE cycle at high $T$ to the low temperature regime with a
  rate close to $1$ complete layer per cycle.

  Our analysis shows that the characteristic transition temperature should depend
  only weakly on the model (or experimental) parameters within a wide range of
  reasonable choices. The key dependence is on the particle flux during Cd and
  Te pulses, respectively, and our model reproduces its effect on the transition
  qualitatively correct. 

  As an attempt towards a more quantitative comparison with
  experiments, one might identify the temperature $T_{res} \approx 0.365$ for 
  a flux of $2 ML/s$ in our model (cf. Fig \ref{tresvsf}) with the
  experimental value $T \approx 290^o C$ by Faschinger et al.\cite{faschinger}
  Thus, the energy scale $\epsilon_d$ may be fixed, and we obtain, for example,
  a value of $E_a \approx 1.13 \, eV$ for the macroscopic activation energy for desorption of Te$^*$
  atoms which roughly agrees with the experimentally determined $0.96 \, eV$.\cite{tatarenko,cibert}
  Furthermore, the value of $T_{res}$ for our simulations with $5 ML/s$ translates into 
  $T_{res} \approx 329^o C$ which agrees well with results of the simple evaporation model
  from Ref. \onlinecite{faschinger}.
  
  We thus believe, that, in principle, our model offers the potential for a quantitative match with
  experimental data. In simpler versions of the model it was possible to 
  reproduce the temperature dependence of the surface reconstruction as well
  as macroscopic sublimation rates on a semiquantitative level.\cite{aachen}
  The larger number of parameters in the current extension, however,
  requires further input from experiment. Moreover, we hope to 
  obtain reliable estimates of microscopic activation barriers from
  first principle quantum chemical calculations as are already available for
  III-VI semiconductors.\cite{grosseprb,grosseprl} Such calculations will be extremely useful
  and should also shed light on the nature of the weakly bound Te$^*$ states on 
  the surface. 
 
  \ \\

  \begin{acknowledgments}
  We would like to thank S. Tatarenko for useful discussions.
  This project was supported by the Deutsche Forschungsgemeinschaft through
  Sonderforschungsbereich 410. 
  \end{acknowledgments}

   \end{document}